# The "...system of constraints"

**Ken Krechmer**, *Senior Member, IEEE*

*Abstract*—This paper proposes that the mathematical relationship between an entropy distribution and its limit offers some new insight into system performance. This relationship is used to quantify variation among the entities of a system, where variation is defined as tolerance, option, specification or implementation variation among the entities of a system. Variation has a significent and increasing impact on communications system performance. This paper introduces means to identify, quantify and reduce such performance variations.

## I. INTRODUCTION

C. Shannon in his seminal work, *The Mathematical Theory of Communications*, describes a communications system as a "...system of constraints..." [1]. In his work, Shannon focuses on the impact of noise on the performance of the communications system. This paper examines the operation of the system of constraints when no noise is present to quantify how variation in the system of constraints impacts performance.

The classic communications system diagram, from Shannon, is shown in Fig. 1. It provides a diagramatic model of a physical communications system. A properly designed and implemented communications system without noise is a system where the probability of the receiver receiving what the transmitter transmits is one. When noise is removed, the ordered nature ($p(a|b) = p(b|a) = 1$) of the system of constraints used for communications is clearer. Understanding the mathematical form of the system of constraints then allows the quantification of what is a properly designed and implemented communications system.

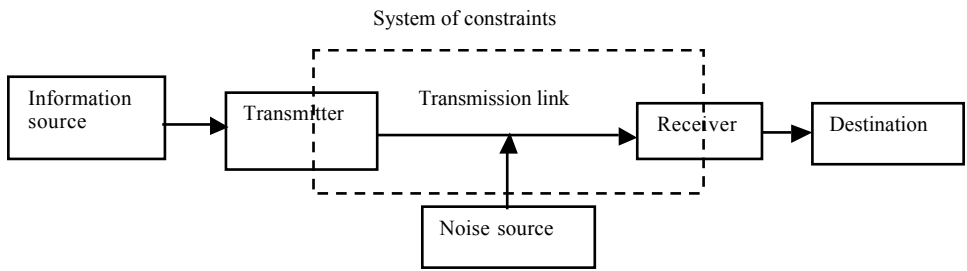

Fig. 1. Physical Model of a Communications system.

Note that the system of constraints is not congruent with the Transmitter, Receiver or Transmission link in this diagram of a physical

Ken Krechmer is a fellow with the International Center of Standards Research, Palo Alto, CA, USA. (phone: 650-856-8836; fax: 650-856-6591; e-mail: krechmer@csrstds.com).

communications system. For this reason a logical model of the functions of a communications system is necessary.

## II. A LOGICAL MODEL OF A COMMUNICATIONS SYSTEM

For communications to occur, any transmitter and receiver must be related by some common reference, whether it be human language (dictionaries and formal syntax provide the reference), ASCII characters, specific frequencies, voltages, or common protocols between a transmitter and receiver. In the last four examples the common reference may be published documents, which may be termed standards.

The philosopher I. Kant first elucidated the idea that a comparison is necessary for any form of understanding [2]. As example, in the course of reading, a word appears of unknown meaning. The reader refers to a dictionary. A dictionary is a common reference that provides the transforms of words into meanings. Assuming that the author also uses a similar dictionary, the reader looks up the unknown word. Upon finding the same word (a comparison), the reader now understands the meaning of the word. This three phase process - apply common reference, compare received signal to reference and identify signal - occurs in any communications process. The three stage process is diagrammatically shown in Fig. 2. The input is applied to the common reference and transformed into a transmitted signal. The received signal is reverse transformed by comparision with the common reference to produce the output. In any communications system a common reference must exist between the transmitter and receiver to create the basis of comparison necessary before communications can occur. In Fig. 1 the common reference appears as the system of constraints. In Fig. 2 the logical model shows the function of the system of constraints specifically.

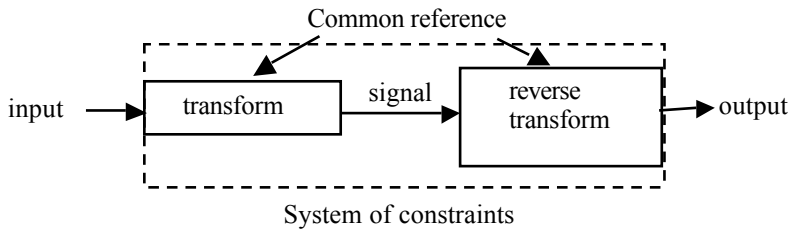

Fig. 2. Logical Model of a Communications System without Noise.

The multiple parameters of the transmitter and receiver that directly relate to each other are the common reference or system of constraints which appear within the dotted line of Fig. 1. Fig. 2 provides a logical diagram showing the relationship between the common reference required for communications and a communications system. Each common parameter of the transmitter and receiver is a common reference, or following Shannon's terminology, a single *set of constraints*. The logical model of the common reference can now be viewed in a set theoretic form.

## III. A MODEL OF A SIMPLE INFORMATION CHANNEL

Fig. 3 (from Abramson [3] plus the dashed rectangle S) presents a set theoretic model of a simple *information channel* consisting of a transmitter of alphabet A with individual elements $a_i$ and total elements t and a receiver of alphabet B with individual elements $b_i$ and total elements r. Alphabet A and alphabet B are related by the existence of a common set of elements S, where S = A ∩ B. Without a common set S (the common reference) no communications is possible. When S = A ∩ B, without noise, and both r and t ≥ n > 0 then S > 0. S = A ∩ B is the set theoretic constraint that defines the common reference in Fig. 3.

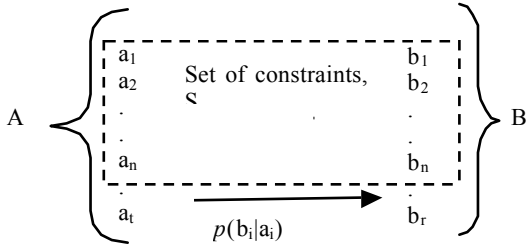

Fig. 3. An information channel.

This model of an information channel represents the transmission link as the probabilistic relationship between ordered pairs of elements. The transmission link's constraints only exist in the choice of the alphabets A and B. The element pairs $a_1$ and $b_1$, $a_2$ and $b_2$, .... $a_n$ and $b_n$ are each defined as a preexisting ordered pair. B. Russell refers to the relationship between each element pair as a one-one relationship [4]. When a transmission link connects one or more ordered pairs communications is possible. Without noise, the sets A and B are related by the ordered relationship of their elements where $p(b_i|a_i)$ equals one, for ordered pairs 1 through n. The set of constraints termed S is formed by the order between the elements of sets A and B and the common alphabet size n. In Fig. 3, the elements $a_{n+1}$ through $a_t$ and $b_{n+1}$ through $b_r$ are not ordered and therefore are not included in the set of constraints.

An example of such an information channel is a human transmitter using the 26 letter English alphabet $a_t$ through $z_t$ and a human receiver using the same alphabet $a_r$ through $z_r$. One condition for error free communications is that the humans use a common alphabet. This condition is a set of constraints consisting of the 26 ordered pairs $a_t$ and $a_r$, $b_t$ and $b_r$, .... $z_t$ and $z_r$. Using this set of constraints the humans are better able to communicate. Without one or more preexisting (before communications) ordered pairs, no reference exists and no comparisons are possible. When the same 26 letter alphabets are used by each person, communications based on the alphabet may occur.

The ordered pairs of a communications system may be created by chemical bonds (A-C, G-T in DNA), preexisting written or spoken alphabets, preexisting word dictionaries or the specifications of the transmitter and receiver (electronic communications) that constrain the implementation of the transmitter, transmission link, and receiver (entities). The definition of an entity here is arbitrary, and indicates some independence from other entities. The simplest entity is a single set (e.g., set A or set B); a complex entity may consist of multiple sets. The preexisting sets of constraints are the system of constraints defining the relationship between two or more entities. In a functioning communications system the implementations of the transmitter, transmission link and receiver are each bound by this system of constraints.

IV. A MODEL OF THE BOUNDS ON A SET

Describing the relationship between the transmitter and the receiver that enables communications first requires a description of each of these entities. In information theoretic terms, the system of constraints defines the bounds of the information channel which includes the transmitter and receiver. Fig. 3 models the simplest possible information channel between two entities as a single set of constraints, S. Considering each entity as a single set, the information theoretic description of the entropy of a single set (A) is:

$$H(A) = -\sum_{i=1}^{i=n} p(a_i) \log p(a_i) \qquad (1)$$

This equation describes the entropy distribution (H) of set A with n discrete random variables $a_i$. The limit of H(A) = log n, which is the bound of the entropy of set A, and H(A) approaches this bound as a limit. The logarithm of the number of elements of the set along with the description of the set (in this case, set A) describes the set bounds in information theoretic terms.

The relationship of an entropy distribution (e.g., the data passing through a transmitter or a receiver) to its bound (common reference) is shown in Fig. 3. This relationship can be explored using the concept of mutual information. Thomas and Cover [5] define mutual information (MI) as:

$$MI = I(A;B) = H(B) - H(B|A) \qquad (2)$$

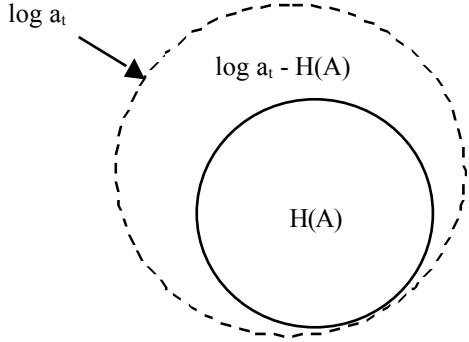

Fig. 4. Venn diagram of a single entropy distribution and its bound.

Fig 4 shows the Venn diagram of a single entropy distribution contained within its bound. In Fig. 4 the entropy distribution is H(A) and its bound (log $a_t$) may be considered H(B) to solve equation 2:

$$H(B|A) = \log a_t - (H(A) \qquad (3)$$

$$MI = \log a_t - (\log a_t - H(A)) \qquad (4)$$

$$MI = I(A, \log a_t) = H(A) \qquad (5)$$

Eq. 5 shows that an entropy distribution when considered in its own context is the mutual information. A similar result is noted in Cover and Thomas [5] (page 20) as *self-information*. However "its own context" may be either the bound of H(A), which must equal the maximum entropy distribution of H(A), or H(A) itself. If the context is not the bound of H(A) or H(A) itself, then there are two or more sets. A single entropy distribution must be considered in the context of its bound (e.g., log $a_i$), as using the context H(A) is self-referential. The author proposes that an entropy distribution and its bound be termed *relative-information*. This is an important point: the information in a set only exists in context and the only logically consistent context is the limit of the entropy distribution.

This change from self-information (referenced to itself) to relative-information (referenced to its bound) does not change the value of information, but it has other ramifications, some of which are developed below.

Fig. 4 is also useful in defining the term *similarity*. By definition all sets that are *similar* to H(A) fall within the limit of H(A). For the purpose of creating a preexisting order (i.e., *similarity*), a specific description of a single entity (which may consist of multiple sets) may be made. Such a description is termed a *similarity* description, as the purpose of making such a description of an entity is almost always to create or maintain similar entities.

## V. A Model of an Information Channel with Bounds

Fig. 1 defines a communications system. Fig. 2 identifies how the transmitter's implementation of the common reference transforms the input into the signal desired for transfer across the channel; the receiver's implementation of the common reference reverse transforms the signal into the output. Fig. 3 describes how an information channel exists within a communications system when the elements of two sets exist as ordered pairs across a transmission link. Fig. 4 develops the relationship of a single entropy distribution to its bound.

Fig. 5 combines the concepts shown in Fig. 3 with the relationships shown in Fig. 4 to model an information channel and its bounds using Venn diagrams. Fig. 5 shows the case where the transmitter (t) and receiver (r) sets each have n ordered pairs. Log $a_n$ is the bound of the transmitted entropy [H(A)] and log $b_n$ is the bound of the received entropy [H(B)]. These bounds are shown as dotted concentric circles around the related entropy H(A) and H(B). Fig. 5 is provided for visualization, not calculation, as the shapes are idealized. The entropy [H(A) and H(B)] always remains within its respective bound (log $a_t$ and log $b_r$).

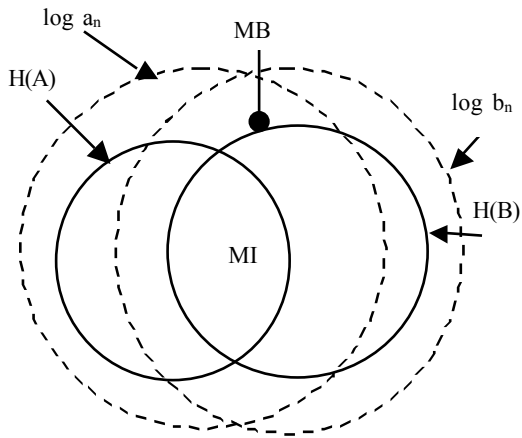

Fig. 5. Venn diagram of an information channel and its bounds.

Fig. 5 models how the relationship between, and bounds on, H(A) and H(B) limit the maximum mutual information. MI (the area within the solid line lens in Fig. 5) is the mutual information transferred across the information channel between the transmitter and the receiver.

The mutual information (MI) transmission equation for the information channel shown in Fig. 5 is:

$$MI = I(A;B) = \sum_{i=1}^{i=n} p(a_i, b_i) \log \frac{p(a_i, b_i)}{p(a_i) p(b_i)} \qquad (6)$$

The mutual information (MI) is the *Kullback Leiber distance* (Cover, 1991) between the joint distribution $[p(a_i, b_i)]$ and the product distribution $[p(a_i)p(b_i)]$. Then log n, the upper limit of I(A; B), is the maximum bound of the information channel. I(A; B) = log n can occur only when the bound of set A for the receiver = log n, and the bound of set B for the transmitter = log n, are overlapping and congruent. I(A; B) = log n only occurs when there is no noise in the communications system.

MB (log $a_n$; log $b_n$), the *mutual bound* (the lens shape enclosed in dotted lines in Fig. 5), is defined as the bound of an information channel. MB is the mathematical form of S shown in Fig. 2. Expanding the equation for MI above into separate joint and product entropy terms:

$$MI = \sum_{i=1}^{i=n} p(a_i, b_i) \log p(a_i, b_i) - \sum_{i=1}^{i=n} p(a_i, b_i) \log p(a_i) p(b_i) \qquad (7)$$

When $p(a_i) = p(b_i) = 1/n$, the bound of set A for the transmitter = log n and the bound of set B for the receiver = log n. This is the mutual bound (MB) of the information channel. The limit of MI, which is MB, may be found by inserting 1/n for $p(a_i)$ and $p(b_i)$. Then the equation for MB is:

$$MB = - (log\ n\ to\ 2\ log\ n) + 2\ log\ n \qquad (8)$$

The product entropy term is 2 log n. The joint entropy term ranges from -(log n to 2 log n) depending upon the *Kullback Leiber distance* [5] which is determined by the noise in a communications system (Fig. 1).

That MB and MI both can be derived from the *Kullback Leiber distance* is an indication that Fig. 5 presents a realistic view of the relationship of the mutual bounds to the mutual information. Using the new concept of mutual bounds it is possible to examine the impact of variation of these bounds on the performance of a communications system.

## VI. MODELING CONSTRAINTS

Consider two sets describing an information channel that has no noise but where the number of elements in each set is different ($a_t \neq b_r$) such as Fig. 3. In this case the difference between the bounds of each of the two sets is the variation caused by unordered elements in sets A and B. Variation is defined as the existence of elements of set A or B that are not ordered pairs. Without variation between set A and set B, S = A ∩ B = log n. Attempting to hold $a_t = b_r = n$, thereby eliminating variation, is the practice in electronic or optical communications system design.

In an operating communications system, the relationship of $a_y$ to $b_r$ to n, for each set of constraints is determined by the actual implementation of each set pair $a_i/b_i$ (Fig. 3). Notice when $a_t$ or $b_r > n$ the information channel is less efficient and this effect is independent of noise. As the communications system implementation approaches optimum, $a_t = b_r = n$ and $p(b_i|a_i) = 1$(no noise), then MB approaches log n as a bound.

The term S (set of constraints), developed above, is the bound of MB. MB describes the bounds of the information channel in the presence of noise while S describes the bounds of an idealized information channel where noise is zero. When the noise is zero, the effect of differences in the bounds of sets A and B on communications system performance may be examined.

Fig. 5 is also useful to describe what is meant by *compatibility*. All sets that have any degree of compatibility with each other have an MB that falls within a bound S. MB = S = log n is the description of the bound on the compatibility of the two sets shown in Fig. 3. A description of a communications system is often made using multiple related sets which creates a specific preexisting order (i.e., compatibility). Such a description is termed a compatibility description. The purpose of making such a description is almost always to create or maintain *compatible* entities.

## VII. Quantifying Variation in a Communications System

In specific designs or implementations of the transmitter and receiver (when the link characteristics are accounted for by the choice of sets A and B), $a_t = b_r = n$ may not be true for each set of constraints in the system of constraints that bound a communications system. When $a_t \neq b_r \neq n$, the design/implementation of the system is less than optimum. A reduction from the optimum is not necessarily undesirable but it should be defined to prevent design or implementation errors. The models developed above assist in evaluating any variation from an optimum communications system design.

Multiple implementations of an actual transmitter or receiver are rarely identical. Fig. 5 shows that differences in similarity directly impact compatibility. Differences in the implementations are caused by differences in the number of elements of a transmitter set ($a_t$), or receiver set ($b_r$) caused by some variation. Such variation (V), which is independent of noise, is caused by errors or misunderstandings in the common reference used (e.g., similarity or compatibility descriptions), errors in the implementations, or the implementation of different options.

The relationship between $a_t$ and $b_r$ may be used to quantify the total variation in incremental parameters. Analog parameters (non-incremental) are usually described using the concept of tolerance which defines a bi-directional variation range. In analog parameters, information variation within the specified tolerance range is ignored; cases where the information variation is beyond the specified tolerance range are considered faults in common engineering practice. For this reason this paper focuses on incremental (non-analog) parameter variation.

In the most efficient communications system, all the receiver states will have a one-one relation with the transmitter states and no additional. This is shown as: log $b_r$ = log $a_t$. In less efficient communications systems, the information variation is V = |log $b_r$ − log $a_t$| for each set of constraints. The sum of the information variation of all the non-ignored non-fault sets of constraints (numbering x) in a communications system is $\Sigma V_i$ for i = 1 to x. As $p(b_i|a_i)$ goes to 1, MI and MB increase. In the simplest communications system, without noise, as V goes to zero, MB goes to log n as a limit, the maximum performance of the communications system.

The information channel shown in Fig. 3 identifies two sets (alphabets) forming the simplest communications channel. Assuming that these alphabets define only one aspect of the coding, other necessary parameters of the transmitter and receiver may include bandwidth, initialization, synchronization, training, framing, error control, compression, session layer protocol, etc. The description of these additional communications parameters entails additional sets of constraints which are each supported across an information channel as described in Fig. 3.

The difference between MB and log n not due to noise is caused by the effects of differing implementations, defined by $\Sigma\ V_x$. V terms could also include the impact of variation related to the design documentation as well as the implementations. Variation may be caused by differences in the similarity of: timer specifications, buffer sizes or revision levels (when the revisions modify the number of elements in any set in the system of constraints); and also by different options, or protocol layers, or revisions that modify the number of elements in any of these at a single end of the communications system.

When multi-protocol layer transmitters and receivers have a variation somewhere in the system of constraints, $\Sigma\ V_x$ will exist as a reduction in the maximum possible MB. Given the current state of design documentation (where each set of constraints is not defined separately), the ability to compare sets of constraints in each protocol layer of a complex communications system to identify possible variation is nearly impossible. And because of the large number of combinations possible, the ability to test all possible combinations of sets of constraints is often close to impossible. Therefore, as communications systems continue to become more changeable and complex, the value of $\Sigma\ V_x$ increases. A new mechanism, which is termed adaptability, has emerged to address this problem and decrease the effective value of $\Sigma\ V_x$.

## VIII. Model of an Adaptable System

Maintaining a common reference in a communications system is usually described as maintaining compatibility between entities. A common reference can also support adaptability between entities. Adaptability, as used here, allows a means to identify, negotiate and select a desired compatible relationship between different, potentially compatible entities. Adaptability is often achieved in conjunction with the operation of humans, that is, not automatically.

Adaptability may be achieved automatically in many ways. An independent common reference may be used with which each entity or protocol set is designed to interoperate. Clark [6], describes the effect of a common reference as interoperation that may be achieved utilizing a spanning layer (e.g. TCP/IP) or common interface standards. In such cases the common reference allows different entities to be used. Another example would be an Edison light bulb socket which supports many different types of lamps. In this example the human user identifies and selects the specific lamp and the Edison light bulb socket (the common reference) makes this adaptation possible. Adaptability may also be created by a common software program that operates as the common reference between both ends of the communications system (one example is agent software). Or a common protocol may be used for the purposes of identification, negotiation and selection. When such a protocol is only used for these purposes it is termed an etiquette [7].

Fig. 6 shows a multi-mode communications system consisting of three independent transmitter and receiver sets and one independent etiquette. Communications is possible using any one set of the compatible transmitters and receivers. The etiquette shown in Fig. 6 is used to negotiate the "best" transmitter and receiver set for a specific communications application. In this simple example, higher S which offers more possible communications states, is considered better.

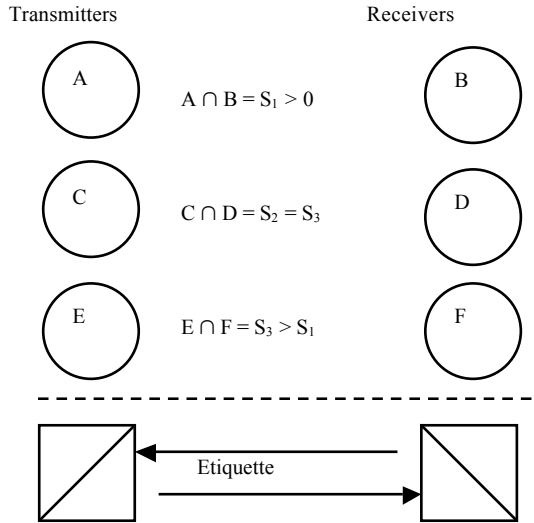

Fig. 6. A multi-mode system with one etiquette.

Consider Fig. 6. Transmitter set A and receiver set B are compatible. Transmitter set C and receiver set D are compatible and equal (in number of states) to transmitter set E and receiver set F. Transmitter set E and receiver set F have more states than A and B. In this example none of the other possible sets are compatible. In this case it is most desirable for the transmitter and receiver selected for operation to be E and F or C and D. Fig. 6, without the etiquette, could also be viewed as a model of a 2G or 3G tri-band cellular mobile and cellular base station. Fig. 6, might also be viewed as a model of a multi-mode software defined radio (SDR).

For the purpose of identifing, negotiating and selecting a common reference for data and control communications, a specific description of the negotiation procedures among multiple possible information channels may be used. Such a description is an *adaptability* description which includes a mechanism to negotiate among multiple possible information channels to achieve the desired compatibility.

## IX.  MINIMIZING VARIATION IN COMMUNICATIONS SYSTEMS

Complex communications systems utilize multiple layers of compatibility standards (e.g., protocols), each of which may exhibit variation. For application to application communications to be efficient, the sum of the total communications system variation ($\Sigma\ V_x$) must be controlled, otherwise MB may be significantly reduced. The $\Sigma\ V_x$ is very difficult to calculate in multi-protocol layer systems with time-independent processes, and testing all possible variations is usually not practical.

Etiquettes can ensure that complex communications systems function properly at the applications layer. Etiquettes define a fully-testable independent protocol (from the data and control layer protocols) whose purpose is to negotiate among the parameters (most or all of the sets of constraints) at the transmitter and receiver to select the common sets known to fulfill the requirements necessary for a specific communications application. The purpose of an etiquette is to support adaptability.

In a communications system, multiple sets (used in multiple OSI layers) exist to define a multi-layered communications interface. Changes to the sets describing the transmitter or receiver or their implementations may create elements that are not contained in the MB of a specific layer or reduce the MB of a different layer (e.g., by changing a buffer size which might reduce maximum packet length). Such changes can cause compatibility problems. When the changes to any compatible sets are a superset of the previous compatibility sets, then $\Sigma$ MB remains constant or increases. However, maintaining a superset in multiple layers of communications protocols is problematic. The ability to create a superset is made practical by requiring that an etiquette be a single tree structured protocol (which may be expanded and will always remain a superset of prior instantiations). An etiquette discovers and then negotiates between multiple transmitter and receiver implementations and their parameters at all required layers of the OSI model (X.200) to identify and select implementations that are most desirable for a required communications application. The etiquette can perform such a negotiation based on knowledge of the desired application, existing compatibility sets or even known "bugs" caused by using specific revisions of the sets of constraints in a desired application.

Etiquettes are already used in many communications systems e.g., ITU G3 fax T.30, ITU telephone modems, ITU V.8, ITU digital subscriber line transceivers G.994.1, IETF Session Initiation Protocol, W3C XML; their properties have been explored by Krechmer [7]. But the value of etiquettes is not widely understood or employed. As example, the 3G cellular standard, IMT-2000, defines five different communications protocols. Currently the means of selecting a specific protocol stack is left to the designer. Existing multimode cellular handsets and base stations sense the strongest signal and give priority to higher generation protocols over lower (a selection mechanism). Such handsets and base stations can support protocol selection, but cannot support protocol negotiation. For a span of time, different protocol stacks will be used in different geographic areas and the negotiation that an etiquette enables is of less value. Eventually however, multi-mode cellular handsets and base stations will appear; then an etiquette becomes more important, not only to negotiate around incompatibilities that emerge as more independent implementations and revisions of the communications standards exist, but also to allow the service provider to select the protocol that optimizes system loading or optimizes geographic coverage, or to allow a user to select the protocol that offers the best economic performance. The use of adaptability mechanisms is a system architecture choice which significantly enhances the long term performance of complex communications systems.

This paper has identified that an entropy distribution (in the vernacular - data) when related to its bound (e.g., log n) is a useful definition of information, and that such information follows a structure based on the mathematical form of the bound. These concepts may have far reaching implications that warrant further exploration.

## X. CONCLUSION

Shannon's theory has pointed the way toward more efficient use of transmission links for over 50 years by identifying the maximum possible data rate for a given level of noise. The approach offered in this paper is that Shannon's theory can also point the way toward more efficient communications specifications and equipment by quantifying the effect of variation on communications systems. Utilizing mechanisms to support adaptability offers the means to control variation in communications systems. Now that some communications designs are closing in on the maximum possible transmission link performance, it is time to address the performance gains and system improvements that can be achieved by controlling variation with adaptability standards.

**Ken Krechmer** (krechmer@csrstds.com) has participated in communications standards development from the mid 1970's to 2000. He actively participated in the development of the International Telecommunications Union Recommendations T.30, V.8, V.8bis, V.32, V.32bis, V.34, V.90, and G.994.1. He was the technical editor of Communications Standards Review and Communications Standards Summary 1990 -2002. In 1995 and 2000 he won first prize at the World Standards Day paper competition. He was Program Chair of the Standards and Innovation in Information Technology (SIIT) conference in 2001 (Boulder, CO) and 2003 (Delft, Netherlands). He is a Fellow at the International Center for Standards Research and a Senior Member of the IEEE. His current activities are focused on research and teaching about standards.